\documentclass[twocolumn,prl,superscriptaddress,showpacs]{revtex4}

\usepackage{graphicx}
\usepackage{times}
\usepackage{amsmath,amssymb}

\begin{document}

\title{Collective States of Interacting Anyons, Edge States, and the Nucleation of Topological Liquids}

\author{Charlotte Gils}
\affiliation{Theoretische Physik, Eidgen\"ossische Technische
Hochschule Z\"urich, 8093 Z\"urich, Switzerland}
\author{Eddy Ardonne}
\affiliation{Nordita, Roslagstullsbacken 23, SE-106 91 Stockholm, Sweden}
\author{Simon Trebst}
\affiliation{Microsoft Research, Station Q, University of California,
Santa Barbara, CA 93106}
\author{Andreas W. W. Ludwig}
\affiliation{Physics Department, University of California, 
Santa Barbara, CA 93106}
\author{Matthias Troyer}
\affiliation{Theoretische Physik, Eidgen\"ossische Technische
Hochschule Z\"urich, 8093 Z\"urich, Switzerland}
\author{Zhenghan Wang}
\affiliation{Microsoft Research, Station Q, University of California,
Santa Barbara, CA 93106}

\date{\today}

\begin{abstract}
Quantum mechanical systems, whose degrees of freedom 
are so-called su(2)$_k$ anyons, form a bridge between ordinary SU(2)  
spin systems and systems of interacting non-Abelian anyons.
Such a connection can be made for arbitrary spin-S systems, and we
explicitly discuss spin-1/2 and spin-1 systems.
Anyonic spin-1/2 chains exhibit a topological protection mechanism 
that stabilizes their gapless ground states and which vanishes only in the 
limit ($k \to \infty$) of 
the ordinary spin-1/2 Heisenberg chain.
For anyonic spin-1 chains we find their phase diagrams to
closely mirror the one of the biquadratic SU(2) spin-1 chain.
Our results describe at the same time nucleation of different
2D topological quantum fluids within a `parent' non-Abelian quantum Hall
state, arising from a macroscopic occupation of localized, interacting anyons.
The edge states between the `nucleated' and the `parent'
liquids are neutral, and correspond precisely to the gapless modes of the anyonic chains.
\end{abstract}


% 05.30.Pr   Fractional statistics systems (anyons, etc.)
% 73.43.Lp  Collective excitations (Hall effects)
% 03.65.Vf   Phases: geometric; dynamic or topological
% 03.67.Lx  Quantum computation
% 73.43.?f   Fractional quantum Hall effect

\maketitle

%%%%%%%%%%%%%%%%%%%%%%%%%%%%%%%%%%%%%%%%%%%%%%%%%%%
% Introduction ---
%%%%%%%%%%%%%%%%%%%%%%%%%%%%%%%%%%%%%%%%%%%%%%%%%%%

\noindent
SU(2) spin degrees of freedom are ubiquitous in condensed
matter physics describing the elementary quantum mechanical 
properties of many magnetic materials.
For ordinary SU(2) spins there is -- mathematically spoken -- 
an infinite number of representations, or in other words arbitrarily 
large spins.
Here we consider a `quantum deformation' of SU(2) 
\cite{SU2q}, where we limit
the number of representations to $k+1$ `angular momenta' 
that take the values $j=0, \frac{1}{2}, 1, \ldots, \frac{k}{2}$.
The degrees of freedom in these so-called su(2)$_k$ 
theories are examples of non-Abelian anyons
with the Ising ($k=2$) 
and Fibonacci ($k=3$) anyons being studied in a variety
of contexts such as unconventional $p_x + i p_y$ 
superconductors \cite{ReadGreen}, fractional
quantum Hall states \cite{ReadRezayiZk},
and 
proposals for topological quantum computation 
\cite{Nayak08}.
We will therefore refer to these generalized angular 
momenta also as anyon types. 
The analog of combining two ordinary spins, and reducing
the tensor product, corresponds to the `fusion' of two anyons
which obey the fusion rules
\begin{equation*}
 j_1 \times j_2 = |j_1-j_2| + (|j_1-j_2|+1) + \ldots + \min(j_1+j_2,k-j_1-j_2) \,.
\end{equation*}
For example, fusing two anyons with generalized angular momenta  $j_{1,2}=\frac{1}{2}$, these
rules imply $\frac{1}{2} \times \frac{1}{2} = 0 + 1$ for $k\geq2$,
which in the limit $k\to\infty$ describes the
coupling of two ordinary spin-1/2's into a singlet or triplet. 
For finite `level' $k$ the
occurrence of several possible fusion outcomes
is the hallmark of what is known as non-Abelian particle statistics.

% short intro to quantum Hall
Here we consider the collective ground state formed by a set of interacting
anyons in the presence of an interaction that energetically splits the possible 
fusion outcomes -- similar to the Heisenberg Hamiltonian for ordinary spins.
In particular, we address the formation of this collective state 
in the context of the original topological liquid of which the anyons are excitations,
such as a non-Abelian quantum Hall liquid. 
Summarizing our results we find that the occupation of a `parent' topolgical
liquid with a finite density of interacting anyons leads to the nucleation of a 
{\em distinct} topological liquid within the original parent liquid.
Specifically, we make an explicit connection between the {\em gapless} 
collective states of chains of interacting anyons and the {\em edge state} 
between these two liquids, which in turn allows us to fully characterize the
nucleated liquid. 
This general correspondence also points to the possible occurrence of 
various unconventional quantum Hall states that arise from interactions
between anyons with generalized angular momentum $j>1/2$.

A first step towards understanding the collective ground states  of interacting 
anyons has  been taken by studying
chains of interacting Fibonacci anyons \cite{GoldenChain,CollectiveStates,Proceedings}:
Uniform chains with pairwise interactions that 
favor either the `singlet' or `triplet'-channel 
are gapless and can be mapped exactly onto the tricritical Ising
and critical 3-state Potts models, respectively 
\cite{GoldenChain}.
In conventional systems these gapless theories would be completely
unstable to the formation of a gap, but a more subtle 
mechanism is at play in the Fibonacci chain where an additional topological 
symmetry stabilizes the gaplessness of the system against local perturbations 
\cite{GoldenChain}.
However, if we consider the more general anyonic systems
described by su(2)$_k$ theories and take the `undeformed'
limit $k \to \infty$, 
we recover a system of ordinary SU(2) spins for which there is no such 
notion of a topological symmetry. 
This observation naturally raises the question whether the observed
topological protection is unique to the Fibonacci theory su(2)$_3$.
In this Letter, we show that the existence of a topological symmetry 
is a common feature in all su(2)$_k$ anyonic theories and that it 
protects the gaplessness
of generalized spin-1/2 chains in these theories
for {\em all finite levels} $k$. 
In fact, it is the ordinary SU(2) spin-1/2 chain that stands out in this series 
as it looses this special symmetry and protection mechanism.
We also discuss su(2)$_k$ generalizations of the spin-1 chain,
a similar topological symmetry protection there, 
and close connections 
between the phase diagram of the
anyonic spin-1 chains and the 
biquadratic SU(2) spin-1 chain.

%%%%%%%%%%%%%%%%%%%%%%%%%%%%%%%%%%%%%%%%%%%%%%%%%%%
% Generalized spin-1/2 chans ---
%%%%%%%%%%%%%%%%%%%%%%%%%%%%%%%%%%%%%%%%%%%%%%%%%%%

\paragraph{Anyonic spin-1/2 chains.--}

We first turn to su(2)$_k$ quantum deformations of ordinary quantum
spin-1/2 chains which have been introduced in Ref.~\onlinecite{GoldenChain}. 
These `golden chains'  consist of a linear arrangement of
non-Abelian anyons with `angular momentum' $\nu=1/2$ in the
su(2)$_k$ theory.
Pairwise interactions between adjacent anyons favor fusion into the 
`singlet' $j=0$ for `antiferromagnetic' (AFM) exchange and 
into the `triplet' $j=1$ for `ferromagnetic'
(FM) exchange.
These su(2)$_k$ spin-$1/2$ chains turn out to be gapless
 for all levels $k$ \cite{GoldenChain}. 
The critical theory is a conformal field theory (CFT)
closely related to the original su(2)$_k$ theory, namely a particular 
coset theory \cite{gko86}.
We first concentrate on AFM couplings for which 
this CFT description is given by the 
$(k-1)$-th unitary minimal model.
For $k=2$ this is the Ising model, for $k=3$ the tricritical Ising model,
and in the limit $k \to \infty$ becomes the $c=1$-theory describing 
the ordinary Heisenberg chain.

For all finite levels $k$, there exists an additional
topological symmetry, that defines $k+1$ symmetry
sectors which correspond to the overall fusion
channel of all anyons in the chain. 
This symmetry corresponds to the operation of commuting 
a spin through all the spins in the chain. While this symmetry
operation also exists in the limit $k\rightarrow\infty$ of the ordinary 
Heisenberg chain, we find that it plays a fundamental role
only in the case of $k$ being finite.
To be explicit, we give
the matrix elements of the topological symmetry
operator $Y$ in terms of the $F$ matrices (which we
define in the auxiliary material \cite{EPAPS}):
\begin{equation}
  \langle x_1',\ldots,x_{L}'| Y | x_1, \ldots, x_{L}\rangle 
  = \prod_{i=1}^{L} \left( F^{x'_{i+1}}_{\nu x_i\nu}\right)^{x_i'}_{x_{i+1}} \,.
  \label{eq:TopologicalSymmetry}
\end{equation}
This topological symmetry becomes important for the anyonic
spin chains as it protects their gapless states against instabilities
arising from local perturbations. 
Since the operator $Y$ commutes with the Hamiltonian, we can
assign each perturbation
to one of its symmetry sectors.
Small, translationally invariant perturbations preserving this symmetry 
can only drive the critical system into a gapped phase if there is an 
operator (at momentum $K=0$), relevant in the
renormalization group sense, which is in 
the same topological sector as the ground state (typically the trivial sector). 
Relevant operators in other 
sectors break the topological symmetry and 
are thus prohibited from opening a gap. 
The number of translationally invariant relevant operators grows as
$k-1$, while the number of topological sectors grows as $k+1$. 
The question thus is whether these two diametrical effects 
result 
in a cancellation, as it is the case for $k=3$, and lead to a topological protection for  $su(2)_k$ 
chains with $k > 3$.

We explain that this topological protection indeed exists for all finite levels $k$
by observing a powerful 
connection between the coset theories describing 
the gapless state
and the assignments of topological symmetry sectors to 
the relevant operators in these theories: 
The primary fields in these coset theories
carry (a pair of) su(2)$_k$ labels like those of the original anyonic degrees of freedom.
This observation allows us to 
obtain topological sectors for all fields
in the gapless theory and identify those operators which for arbitrary $k$
can drive the system into a gapped phase.
In the limit $k\rightarrow\infty$ we recover the 
behavior of the ordinary spin-1/2 chain.
We have checked, for the levels $k=2,3,4,5$, that the so-obtained topological 
assignments agree with results from exact diagonalization of chains with
both AFM and FM interactions with up to $L=24$ anyons (for $k=5$)  
using Eq.~(\ref{eq:TopologicalSymmetry}).

%-----------------------------
\begin{figure*}[t]
\includegraphics[width=\linewidth]{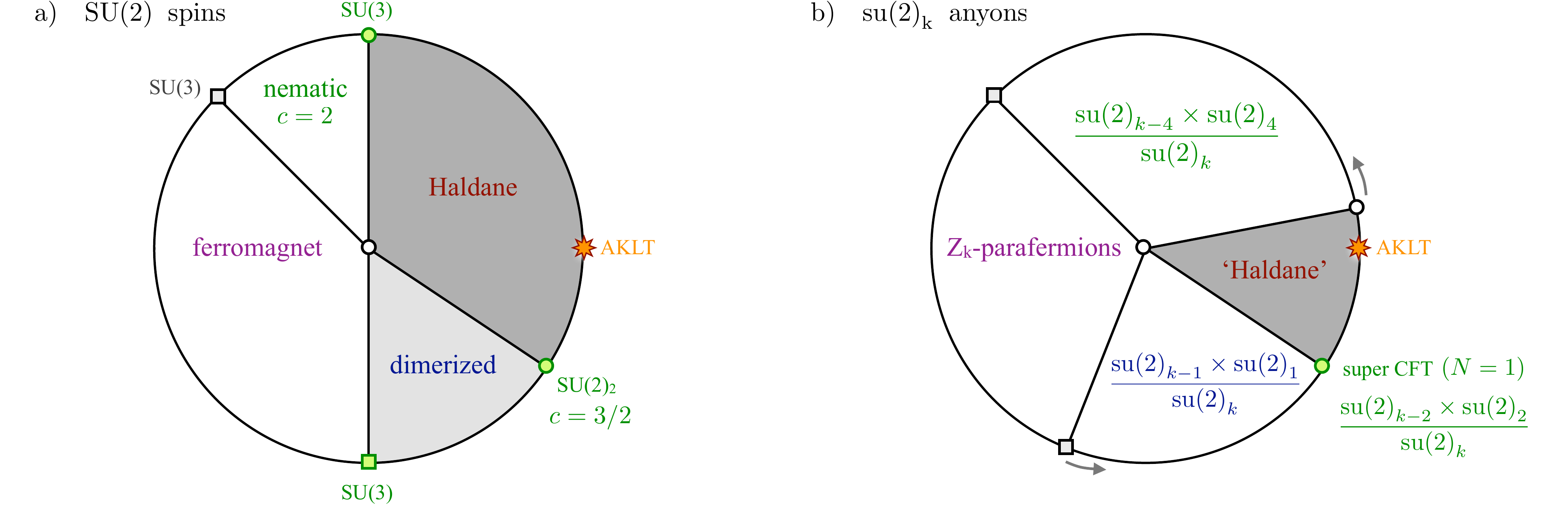}
\caption{
   (color online)
   Phase diagrams of the `biquadratic' spin-1 chain for a) ordinary SU(2) spin-1's 
   and b) the anyonic  su(2)$_k$-theories with $k \geq 5$.
   Projections onto the triplet ($j=1$) and quintuplet ($j=2$) states
   are parametrized by an angle $\theta$ as $J_{\rm 1}=-\sin(\theta)$ and $J_{\rm 2}=\cos(\theta)$.
 }
\label{Fig:PhaseDiagramSpin1}
\end{figure*}
%-----------------------------

In describing the details of the above topological symmetry assignments
we will concentrate
for brevity
 on the case that $k$ is odd. 
In this case we can restrict ourselves to the `integer spin' representations
$0,1,\ldots,(k-1)/2$ of su(2)$_k$. The generalized spin-1/2 chains are 
then based on anyons carrying `angular momentum' $j=(k-1)/2$ 
\cite{alphachain}.
The fields $\phi^{j_1}_{j_2}$ in the gapless theories of 
the AFM  chains carry two labels $j_1, j_2$.
In the coset construction of these theories, 
i.e. ${\rm su(2)}_{k-1} \times {\rm su(2)}_{1}/{\rm su(2)}_k$,
the label $j_2$ corresponds to the representations of su(2)$_{k-1}$,
the label $i_2$ to those of su(2)$_{1}$,
and
$j_1$ to those of su(2)$_k$. 
The label $j_1$ turns out to determine the topological sector of the fields.
In particular, the fields in the topologically trivial sector
turn out to be $\phi^{0}_{j_2}$.
We consider the `character decomposition'
$\chi^{(1)}_{i_2}\chi^{(k-1)}_{j_2}=\sum_{j_1}B^{j_1}_{j_2} \chi^{(k)}_{j_1}$, 
where
$\chi^{(k)}_j$ denotes
the (`affine') character of su(2)$_k$
(and $i_2=j_1-j_2\bmod 1$),
and the $B^{j_1}_{j_2}$'s 
the (Virasoro) characters~\cite{FieldIdentificationsInWords}
of a unitary 
minimal
model~\cite{gko86}.
The $Z_2$ symmetry of these coset models also allows us to identify the sublattice
symmetry of the primary fields. The states at $K=0$ correspond to fields with integer $j_2$, 
while the states at $K=\pi$ correspond to fields with half-integer $j_2$.
The scaling dimensions  which result from this decomposition
are  those of the `Kac-table', 
$x(k,j_1,j_2) = 2\{1 +  j_2(j_2+1)/(k+1) - j_1(j_1 +1)/(k+2)\}$.
It easily follows that the ground state at $K=0$ is in the trivial topological sector.
The lowest lying field at $K=0$ in this sector is $\phi^0_1$, with scaling dimension 
$2(k+3)/(k+1)$. Thus, we find that for any finite $k$, there is no relevant operator 
(with scaling dimension $<2$) which can drive the system into a gapped phase.
In the limit $k\rightarrow\infty$ of the ordinary spin-1/2 chain, this operator
becomes exactly marginal. 
At momentum $K=\pi$, the most relevant field in the trivial sector is $\phi^{0}_{1/2}$, 
with scaling dimension $\frac{1}{2}(k+4)/(k+1)$ which is bounded
by one for $k\geq 2$: staggered perturbations 
can always drive the system into a gapped 
`dimerized' phase. 
We note that the results above hold for $k$ even as well; 
with the difference that
all $k+1$ topological sectors are present.

The critical behavior for FM interactions is described by the $Z_k$-parafermion coset
${\rm su(2)}_k/{\rm u}(1)$, whose fields also carry two labels, 
$\psi^{j}_{m}$. 
Again $j=0,1,\ldots,(k-1)/2$ (for $k$ odd) originates from su(2)$_k$ and determines
the topological sector as before, and $m=0,1,\ldots,k-1$ determines the momentum, 
$K=2 \pi m/k$. 
The fields in the trivial sector are the $Z_k$-parafermion fields, $\psi^{0}_{m}$, of 
scaling dimensions $2 m(k-m)/k$. 
The only state whose corresponding scaling dimension is
less than two at momentum $K=0$ in the trivial sector is the 
ground state, e.g. the identity $\psi^{0}_{0}$, implying that for all $k$ this critical phase is  
stable against small perturbations which do not break translational invariance.

%%%%%%%%%%%%%%%%%%%%%%%%%%%%%%%%%%%%%%%%%%%%%%%%%%%
% Spin-1 chains ---
%%%%%%%%%%%%%%%%%%%%%%%%%%%%%%%%%%%%%%%%%%%%%%%%%%%

\paragraph{Anyonic spin-1 chains.--}

We now turn to  
chains formed by a set of anyons with generalized `angular momentum' $\nu=1$
in the su(2)$_k$ theory for $k \geq 4$. There are three fusion channels of
two $\nu=1$ anyons, namely $1\times 1 = 0 + 1 + 2$, so the most general Hamiltonian 
can be written in terms of an angle $\theta$ as
${\mathcal H}_{\nu=1}  = \sum_{i} \cos\theta \, \Pi^2_{i} -\sin\theta \, \Pi^1_{i}$, 
where the projectors $\Pi^1_{i}$ and $\Pi^2_{i}$ assign an energy $+1$ 
for anyons at sites $i$ and $i+1$ fusing into $1$ or $2$, respectively
(see also \cite{EPAPS}).
We recover the ordinary biquadratic SU(2) spin-1 chain in the limit $k \to \infty$ 
which has a rich phase diagram: 
There are two gapped phases, the Haldane phase~\cite{HaldaneConjecture} and
a dimerized phase \cite{DimerizedPhase}, and two gapless phases, a ferromagnetic 
and a nematic phase \cite{NematicPhase}.

We have calculated the phase diagrams of anyonic 
spin-1 chains for the su(2)$_5$ and su(2)$_7$ theories using exact diagonalization, 
which allows us to generalize this phase diagram to all levels $k \geq 5$.
As illustrated in Fig.~\ref{Fig:PhaseDiagramSpin1}b),
there are four phases that {closely mirror their} $k \to \infty$ counterparts: 
The Haldane phase survives as a gapped phase
and includes a generalized AKLT point where the exact ground states are known
\cite{LongPaper}. 
It is surrounded by two gapless phases which have the same sublattice symmetries as their SU(2) 
counterparts, e.g. a $Z_3$ sublattice symmetry for the phase in the upper wedge
which becomes the nematic phase \cite{FathPeriodTripling}, 
and a $Z_2$ sublattice symmetry for the phase in the lower wedge 
which turns into the dimerized phase.
For arbitrary $k \geq 5$ the gapless phase in the upper wedge is the $k$-th member of  
the family of coset models ${\rm su(2)}_{k-4} \times {\rm su(2)}_{4} / {\rm su(2)}_k$.
For $k=5,7$ we have numerically verified that the low-energy spectra indeed 
match these theories of central charges $c^{(4)}_k=$ $2-24/(k^2-4)$.
Similarly, our numerical results suggest that the gapless phase in the lower wedge is 
for arbitrary $k \geq 5$ the $k$-th member of the family of the so-called off-diagonal 
modular invariants~\cite{Cappelli} of the unitary minimal models ${\cal M}_k$.
The critical endpoints of this phase with the Haldane phase correspond
to the $N=1$ supersymmetric minimal models with central charge $c=3/2-12/[k(k+2)]$.
The chain Hamiltonian at this point can be mapped to an exactly integrable model
\cite{IntegrableModel}. 
In the $k \to \infty$ limit this critical point turns into the su(2)$_2$ Wess-Zumino-Witten (WZW) 
point of the ordinary biquadratic spin-1 chain. 
As in the spin-1/2 case, all extended critical phases are protected by a topological symmetry.

%%%%%%%%%%%%%%%%%%%%%%%%%%%%%%%%%%%%%%%%%%%%%%%%%%%
% Quantum Hall liquids ---
%%%%%%%%%%%%%%%%%%%%%%%%%%%%%%%%%%%%%%%%%%%%%%%%%%%

\noindent
\paragraph{Quantum Hall Liquids.--}
Our program of exploring collective states
of anyonic spin chains is, at the same time, 
a tool to systematically study 
topological
phases which can occur inside  non-Abelian 
quantum Hall liquids due to 
population of such liquids by a macroscopic number
of interacting
non-Abelian anyons.
This also provides us with
the properties of the (so far unexplored) edge states appearing
at the interfaces between these two liquids.
Let us focus for brevity
on {\it bosonic} quantum Hall fluids
\cite{RotatingBoseCondensates}.
Corresponding statements for {\it fermionic} states with the same
non-Abelian statistics \cite{ReadRezayiZk}
involve only differences in (trivial) Abelian factors.

\begin{figure}[t]
\includegraphics[width=\columnwidth]{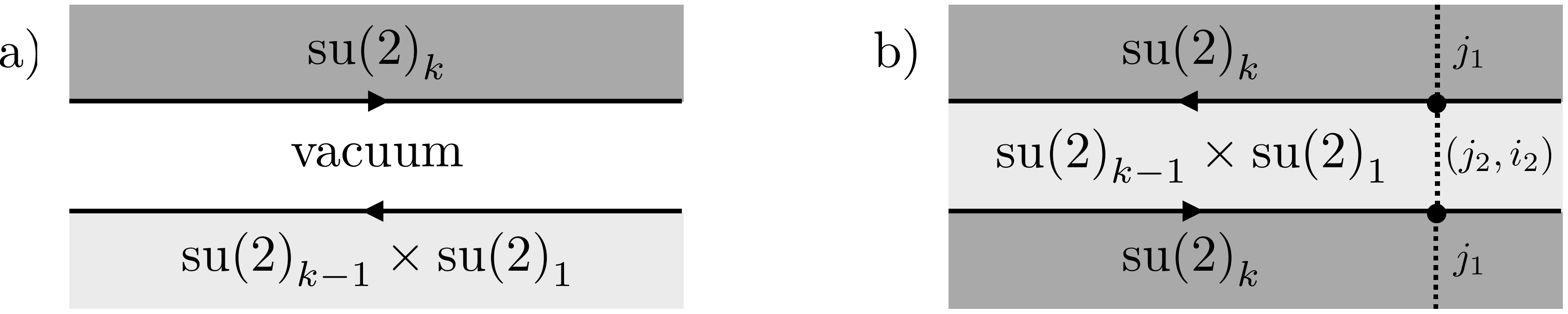}
\caption{
   Topological liquids and their edge states. }
\label{fig:Liquids2}
\end{figure}

First reconsider the case of a linear arrangement of $j=1/2$ anyons in a 
surrounding su(2)$_k$ topological 2D fluid
with pairwise 
interactions favoring singlet pairs. 
As we have seen, the  
gapless state is the minimal model ${\cal M}_k$.
We can think of this critical state as two non-interacting counterpropagating
(neutral) edge states of central charge $c_k =1-6/(k+1)(k+2)$,
which are basically located `on top of each other',
residing in the  surrounding su(2)$_k$ topological bulk fluid.
Let us now imagine separating these two edge states slightly in space,
so that a narrow strip opens between them, see Fig.~\ref{fig:Liquids2}b).
Let us ask
if it  is possible to place {\em another} topological quantum Hall fluid $X$
into this narrow strip such that the above 
pair of counterpropagating  edges of central charge $c_k$
are precisely the edge states between the surrounding topological su(2)$_k$
fluid and the new, intervening 
fluid $X$ in the strip.
The answer to this question is `yes':
The intervening liquid $X$ is a topological fluid 
characterized
by ${\rm su(2)}_{k-1}\times{\rm su(2)}_1$.
This can be seen \cite{Footnote} from the coset representation~\cite{gko86} of the
minimal model
${\cal M}_k=$
${\rm su(2)}_{k-1}\times {\rm su(2)}_1/{\rm su(2)}_{k}$.
Recall that the central charges of 
$X=$ ${\rm su(2)}_{k-1}\times {\rm su(2)}_1$ (numerator) and
the surrounding fluid ${\rm su(2)}_k$ (denominator) 
differ by $c_k$, thus resulting in an edge state between the two liquids
of central charge $c_k$ \cite{Semion}.  
To summarize, our results say that the (AFM) interactions between
an array of $j=1/2$ anyons in an ${\rm su(2)}_k$ topological liquid
nucleate a new intervening liquid characterized by
the topological properties of ${\rm su(2)}_{k-1}\times {\rm su(2)}_1$
\cite{FootnoteTwoSpeciesLiquid}.

Even though our anyons were initially confined to one dimension, 
these results will also apply to a macroscopic number of interacting
non-Abelian anyons occupying two-dimensional (2D)
regions of the surrounding liquid, thereby nucleating larger,
2D regions of the intervening liquid.
The simplest example of this phenomenon was observed 
 \cite{ReadLudwigAbsenceMetal}
for the Moore-Read Pfaffian
quantum Hall liquid,
which, when occupied 
with a macroscopic number of interacting
quasi-hole anyons (even at random positions),  
turns into the simple Abelian `strong-pairing' state. 
In general, the CFT of the edge state 
between the nucleated and the surrounding `parent' liquid
is given by the coset construction.
For the simpler case of FM anyon interactions
the liquid $X$ is Abelian,  and a $Z_k$ parafermion 
CFT edge replaces the neutral  ${\cal M}_k$ edge CFT.
The Pfaffian case above corresponds to $k=2$.

%-----------------------------
%-----------------------------

Even more interesting examples arise when we express our results
for  the anyonic spin-1 chains in terms of intervening liquids.
% upper wedge
A family of novel liquids with topological character~\cite{CommentOnlyAbelianSU2Four}
${\rm su(2)}_{k-4}\times {\rm su(2)}_4$
nucleated within a ${\rm su(2)}_k$ 
surrounding liquid
corresponds to the family of critical phases in the upper wedge 
of Fig.~\ref{Fig:PhaseDiagramSpin1}~b)
with corresponding edge states 
described by the coset 
${\rm su(2)}_{k-4}\times {\rm su(2)}_4/{\rm su(2)}_{k}$.
% lower wedge
Interestingly, the critical phases in the lower wedge 
of Fig.~\ref{Fig:PhaseDiagramSpin1}~b)
describe the appearance of edge states 
between the ${\rm su(2)}_k$ bulk fluid and a novel topological quantum liquid
of type
${\rm su(2)}_{k-1}\times {\rm su(2)}_1$,
with the property that the pair of edge states represents
the {\it off-diagonal} modular
invariant~\cite{Cappelli} of the  minimal model ${\cal M}_k$.
In all cases, the new resulting liquid has {\em less} anyon 
types than its ${\rm su(2)}_k$ `parent' liquid,  due to
the macroscopic occupation
of the latter liquid by interacting anyons.
The nucleation process thus amounts to a reduction of the 
non-Abelian statistics.
Recent work in \cite{Schoutens} also discusses the coset description of an interface between two
different non-Abelian quantum liquids, one spin-polarized and the other spin-singlet.
There, the reduction of non-Abelian statistics originates from
polarization of spins.

In the context of topological liquids the topological symmetry described 
earlier acquires a very physical meaning:
Local perturbations of the chain Hamiltonians correspond to tunneling
events across the intervening liquid. 
It is precisely the perturbations in the topologically trivial sector
which correspond to tunneling processes that are not
accompanied by simultaneous ejection into the surrounding
liquid of anyon particles with non-trivial topological quantum numbers
[such processes correspond to $j_1=0$ in Fig \ref{fig:Liquids2}b)]. 

%%%%%%%%%%%%%%%%%%%%%%%%%%%%%%%%%%%%%%%%%%%%%%%%%%%
% Acknowledgments ---
%%%%%%%%%%%%%%%%%%%%%%%%%%%%%%%%%%%%%%%%%%%%%%%%%%%

We thank D.~A.~Huse for collaboration on a closely related project
and M.~Freedman,  A.~Kitaev, M.~P.~A.~Fisher, A. L\"auchli and K.~Schoutens for discussions. 
Our numerical work employed the ALPS libraries \cite{ALPS}.  
A.~W.~W.~L. was supported, in part, by NSF DMR-0706140.

{\em Note added.--} After the submission of this manuscript, a closely related paper on edges
between different non-abelian quantum Hall states appeared \cite{Bais}. 

%%%%%%%%%%%%%%%%%%%%%%%%%%%%%%%%%%%%%%%%%%%%%%%%%%%
% References ---
%%%%%%%%%%%%%%%%%%%%%%%%%%%%%%%%%%%%%%%%%%%%%%%%%%%


\begin{thebibliography}{99}
\bibitem{alphachain}
The fusion rules imply that $1/2\times k/2 = (k-1)/2$
and $(k-1)/2 \times (k-1)/2 = 0 + 1$. The spin-1/2 chains we
consider for odd $k$ are based on the
$j=(k-1)/2$ representation.
\bibitem{Semion}
We can also rewrite this statement by bringing the Abelian factor ${\rm su(2)}_1$,
a semion, from the fluid $X$ to the edge. We then arrive
at the statement that juxtaposition of the two topological fluids
${\rm su(2)}_k$ and ${\rm su(2)}_{k-1}$, namely Read-Rezayi states,
along a line yields an edge consisting of the (neutral)
${\cal M}_k$ theory and a charged semion $su(2)_1$ moving in the opposite
direction.
For corresponding  fermionic quantum Hall fluids
the edge state between parafermionic $Z_k$ and $Z_{k-1}$
Read-Rezayi states is the neutral CFT ${\cal M}_k$ and an oppositely moving
Abelian $u(1)$ edge state. For $k=3$, a tricritical Ising model ${\cal M}_3$
neutral edge appears between the $Z_3$ Read-Rezayi and $Z_2$ Moore-Read
states.
\end{thebibliography}
\end{document}